# Near-zero-sidelobe optical subwavelength asymmetric focusing lens with dual-layer metasurfaces

Jianlan Xie[1†], Shengnan Liang[2†], Jianjun Liu[1]*, Pinghua Tang[2]*, Shuangchun Wen[1]

*[1]Key Laboratory for Micro/Nano Optoelectronic Devices of Ministry of Education & Hunan Provincial Key Laboratory of Low-Dimensional Structural Physics and Devices, School of Physics and Electronics, Hunan University, Changsha 410082, China*

*[2]School of Physics and Optoelectronics, Xiangtan University, Xiangtan 411105, China*

[†]These authors contributed equally to this work.

**\*E-mail: jianjun.liu@hnu.edu.cn, pinghuatang@xtu.edu.cn**

**Abstract:** The existing metasurfaces with ultrathin volume for asymmetric transmission were often constructed by metal with low efficiency in optical frequency, and could not realize the optical asymmetric transmission and focusing simultaneously. Although the acoustic asymmetric focusing in asymmetric focusing lens (AFL) was realized, the obtained focal point was accompanied by large sidelobe. To solve those problems, an AFL with dual-layer metasurfaces designed by using dielectric materials can realize optical asymmetric transmission and focusing simultaneously in this paper. Furthely, through optimizing the design theory of AFL, the near-zero-sidelobe (NZS) focusing can be realized on the subwavelength scale. The NZS asymmetric focusing of the optimized AFL is effective in broadband, which is expected to possess tremendous potential in the optical micro-nano processing,

## 1. Introduction

In recent years, with the in-depth research on metamaterial, two-dimensional metamaterial with the same electromagnetic properties and designable characteristic—metasurface [1–3] was proposed, which rapidly became the research highlight owing to it controls light based on its subwavelength cell structures and can greatly improve the sensitivity of device performance. The mechanism of wavefront shaping by metasurface is controlling the phase [4], amplitude [5], and polarization [6] of the incident light wave. The optical characteristics of metasurface depend on the geometries and arrangements of its subwavelength cell structures, which has the superiority of controlling light waves. Recently, more and more attention was paid to its easy preparation and low insertion loss. In the fields of super-resolution focusing and imaging, the metasurface has become the mainstream [7–10] by replacing the photonic crystal [11–18] and traditional metamaterial [19] owing to its characteristics of ultrathin volume and high image quality. Meanwhile, various optical devices based on the metasurface have been widely proposed, including vortex beam generator [20], holography device [21], beam deflector [22], and so on.

Asymmetric transmission [23,24] has attracted widespread attention owing to its potential value in integrated optics. In the past two decades, the field of asymmetric transmission has obtained important developments due to the rise of artificial microstructure materials, including the one-way transmission of light waves [25,26], unidirectional invisible [27,28], and unidirectional polarization surface plasmon [29,30]. Based on these characteristics, the micro photon devices such as optical isolators [31], unidirectional polarizers [32], and photodiodes [33–35] have been developed. Artificial microstructure materials, with their novel optical properties and flexible controllability, created conditions for asymmetric transmission of light. However, the above-mentioned artificial microstructure materials could not be widely used due to the limitation of the volume, that is, their size was normally much larger than the operating wavelength. On the contrary, the metasurface, which could adjust the phase wavefront within the subwavelength thickness resulting from its controlling effect of phase wavefront was far greater than the phase accumulation effect of the traditional artificial microstructure materials, has become an ideal choice to achieve asymmetric transmission [35–38]. However, the existing metasurfaces for asymmetric transmission were often constructed by metal [35,36]. Thus, large material loss was unavoidable in the optical frequency region, which caused asymmetric transmission with low efficiency. At the same time, the existing optical metasurfaces could only realize the optical asymmetric transmission [35–38], it could not realize the optical asymmetric transmission and focusing simultaneously. The simultaneous control of multiple light waves can be widely used in the fields of optical micromachining [39], quantum communication [40,41], optical micromanipulation [42,43], and microimaging [44,45], so it has a higher application value than that of single optical asymmetric transmission. Researchers have realized acoustic asymmetric focusing by using AFL [46]. Nevertheless, the obtained focal point was accompanied by large sidelobe, that is, the sidelobe intensity exceeded half of the intensity of the focusing. Therefore, this type of AFL could not be applied in the fields that need high quality of focusing such as acoustic micromachining, micromanipulation, and ultrasonic imaging.

In this paper, the dual-layer metasurfaces are designed by using dielectric materials based on the generalized Snell's law to realize the asymmetric transmission. Then, to expand the application scope of the dual-layer metasurfaces, an AFL with dual-layer metasurfaces is constructed by making phase delays of metasurfaces interacted with each other. When light waves are incident from the left side, they can focus on the right side of the lens, but when light waves are incident from the right side, they cannot arrive at the left side of the lens. Finally, considering the drawbacks of the middle region of metasurface on the right side cannot be involved in focusing, as well as the sidelobe of optical asymmetric focusing is large, the design theory of AFL is optimized further, which can realize the NZS asymmetric focusing on the subwavelength scale. The NZS asymmetric focusing of the optimized AFL is effective in broadband, which has great application potential.

**2. Model and theory**

Based on the generalized Snell's law [4], when the background is air, the refractive angle when the plane wave passes the metasurface can be expressed as

$$\sin\theta_t = \sin\theta_i + \frac{1}{k}\frac{d\varphi}{dy} \qquad (1)$$

where $\theta_i$ and $\theta_t$ are the incident and refractive angles respectively, $\varphi$ is the phase delay generated by the metasurface, $k$ is the wavenumber in the air. It is shown that in Eq. (1), through properly designing the phase delay of the metasurface, the arbitrary refractive angle can be achieved. The cell structure of the metasurface is composed of two kinds of mediums with different refractive indexes. The phase delay is controlled by adjusting the ratio of the two mediums in the cell structure, which can control the refraction of light waves as shown in Fig. 1.

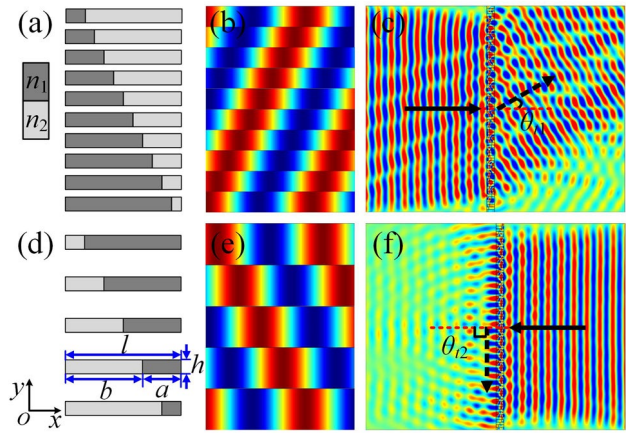

FIG. 1. (a)(d) The cell structure of the metasurface; (b)(e) The intensity distribution of light waves in different cell structures; (c)(f) The refraction and surface wave of light waves are realized by metasurface.

To control the refraction of light waves, the length and height of the cell structure remain constant, that is, $h = 0.2\lambda$, and $l = 0.6\lambda$, where $\lambda$ is the incident wavelength, whose initial value is set to 5 μm, corresponding to 60 THz in air. The background of the metasurface is air. The materials of the cell structure are silicon ($n_1 = 3.42$) and silica ($n_2 = 1.42$), with the corresponding lengths of $a$ and $b$, respectively, and $a + b =$

$l$. The phase delay is controlled by adjusting the values of $a$ and $b$. The transmission characteristics and focusing of light waves are analyzed by the finite element method, and the light waves here we studied are all TM mode.

If the incident angle $\theta_i$ of the plane wave is 0, Eq. (1) can be expressed as

$$\sin\theta_t = \frac{1}{k}\frac{d\varphi}{dy} \qquad (2)$$

Meanwhile, according to the relationship of the refractive index and phase delay, the phase delay caused by the cell structure of the metasurface becomes

$$\varphi = n_1 ka + n_2 kb \qquad (3)$$

By binding of Eqs. (2) and (3), we can obtain

$$\sin\theta_t = \frac{n_1\Delta a + n_2\Delta b}{h} \qquad (4)$$

where $\Delta a$ and $\Delta b$ are the length differences of the same refractive index parts of two adjacent cell structures. Meanwhile, owing to $a + b = l$, $\Delta a = -\Delta b$. Therefore, Eq. (4) can be simplified to

$$\sin\theta_t = \frac{(n_1 - n_2)\cdot\Delta l}{h} \qquad (5)$$

where $\Delta l = \Delta a = -\Delta b$. Thus, the control of the refractive angle can be realized by designing $\Delta l$. The values of $\Delta l$ in Figs. 1(a) and 1(b) are set to $0.05\lambda$ and $-0.10\lambda$, respectively. The intensity distributions of light waves in each cell structure of the metasurface in Figs. 1(a) and 1(d) are shown in Figs. 1(b) and 1(e), respectively. The phase delay of light waves covers the whole $2\pi$ range, which means that the refraction of light waves can be controlled by the metasurface. Meanwhile, according to the Eq. (5), the refractive angle $\theta_{t1} = 30°$ (here we define the refractive angles are positive and negative when their directions are upward and downward, respectively) and $\theta_{t2} = -90°$ are realized by the metasurfaces corresponding to the Figs. 1(b) and 1(e). These results are shown in Figs. 1(c) and 1(f), which indicates that the arbitrary refractive angle even surface wave can be realized by designing the phase delay of the metasurface.

## 3. Results and discussion

### *3.1 The asymmetric transmission realized by the dual-layer metasurfaces*

According to the generalized Snell's law, the refractive light waves are no longer formed instead of becoming the surface wave that propagates along the surface when $\theta_i = 0°$ and $|\theta_t| = 90°$, where $\theta_i$ is the incident angle, $\theta_t$ is the refractive angle of light waves which caused by the metasurface. Moreover, if $|\theta_t| = 90°$ when $\theta_i = 0°$, then when the light waves are incident with certain angles and their signs are opposite to $\theta_t$, they can pass the metasurface. Based on the above theory and the metasurfaces in Fig. 1, if $\theta_{t2} = -90°$, then when $\theta_{t1} > 0°$, the asymmetric transmission can be realized by the dual-layer metasurfaces, as shown in Fig. 2.

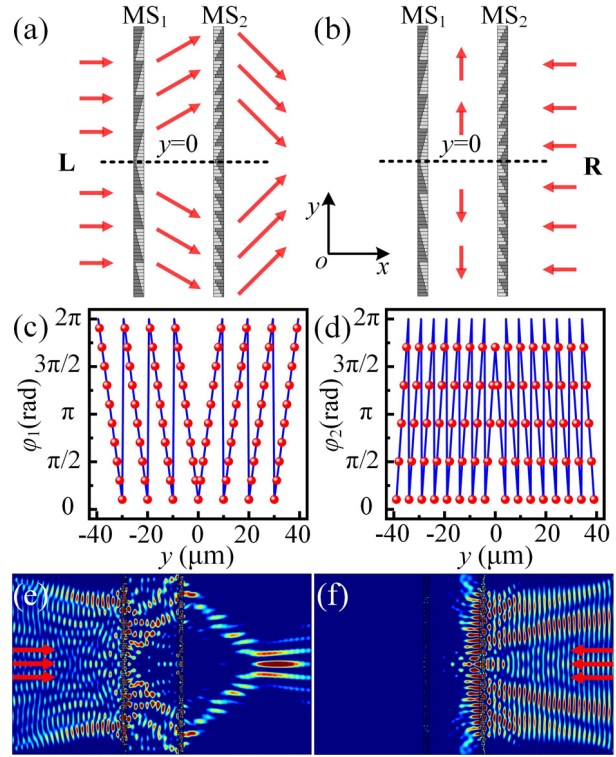

FIG. 2. The principle sketch maps of asymmetric transmission which realized by the dual-layer metasurfaces: (a) light waves can arrive at the right side when light waves are incident from the left side; (b) light waves cannot arrive at the left side when light waves are incident from the right side; Distribution of phase delay: (c) $MS_1$; (d) $MS_2$; Asymmetric transmission: (e) the field distribution when light waves are incident from the left side; (f) the field distribution when light waves are incident from the right side.

As shown in Fig. 2(a) (light waves are incident from the left side) and Fig. 2(b) (light waves are incident from the right side), the dual-layer metasurfaces are $MS_1$ and $MS_2$, whose parameters correspond to Figs. 1(a) and 1(d), respectively. The distributions of the phase delays of these two metasurfaces are symmetric about the $y = 0$. The phase delays $\varphi_1$ and $\varphi_2$ which respectively belong to $MS_1$ and $MS_2$ can be obtained by Eqs. (2) and (3), as shown in Figs. 2(c) and 2(d), in which the blue solid line is the phase delays calculated by theory and the red balls are the phase delays of 79 cell structures in $MS_1$ and $MS_2$. Owing to the distribution of phase delay in $MS_1$ is symmetric about $y = 0$ and the parameters of the structure are consistent with Fig. 1(a), $\Delta l$ of upper and lower parts are $0.05\lambda$ and $-0.05\lambda$, respectively. Thus, according to Eq. (5), the refractive angles of the upper and lower parts are $\theta_{tl1U} = 30°$ and $\theta_{tl1D} = -30°$, respectively, when light waves are normally incident on $MS_1$. Meanwhile, due to $MS_1$ and $MS_2$ are placed in parallel, the refractive angle caused by $MS_1$ is equal to the incident angle in $MS_2$, that is, $\theta_{il2U} = \theta_{tl1U}$ and $\theta_{il2D} = \theta_{tl1D}$. Similarly, $\Delta l$ of upper and lower parts in $MS_2$ are $-0.10\lambda$ and $0.10\lambda$, respectively. Thus, the refractive angles of the upper and lower parts when light waves are normally incident on $MS_2$ are $\theta_{tl1U} = -90°$ and $\theta_{tl1D} = 90°$, respectively, which means that the surface wave is realized. At the same time, since $\theta_{il2U}$ and $\theta_{il2D}$ are contrary to $\theta_{tr2U}$ and $\theta_{tr2D}$, respectively, the light waves can pass the $MS_2$ when they are incident from the left side. Therefore, the asymmetric transmission of light waves can be realized by the dual-layer metasurfaces. The corresponding simulation results of the field distribution are shown in Figs. 2(e) and 2(f).

### *3.2 The asymmetric focusing realized by AFL*

Although the asymmetric transmission can be realized by the dual-layer metasurfaces as shown in Fig. 2, the single function limits its application in the more complex light field. To solve this problem, the structure of $MS_2$ is kept the same as Fig. 2(a), but the structure of $MS_1$ is changed by associating with the phase delay of

MS$_2$ (the specific theoretical derivation and explanation can be seen below), which can realize the more complex asymmetric focusing. Its principle sketch maps are shown in Fig. 3.

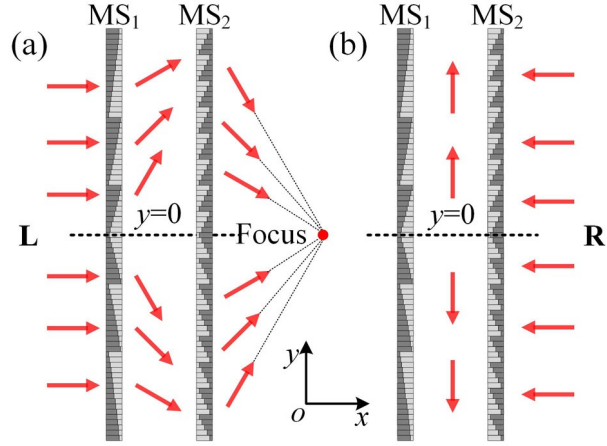

FIG. 3. The principle sketch maps of asymmetric focusing which realized by AFL: (a) the focusing of light waves can be realized when they are incident from the left side; (b) light waves cannot arrive at the left side when they are incident from the right side.

As shown in Fig. 3(a), to realize focusing when light waves are incident normally from the left side, the refractive angle caused by MS$_2$ needs to maintain the distribution of focusing, that is $|\theta_{tl2}|$ varies directly as the distance from the *x*-axis. To realize the goal, the structure of MS$_2$ is kept the same with the MS$_2$ in Fig. 2(a), but the structure of MS$_1$ is changed. Owing to MS$_1$ and MS$_2$ are placed in parallel, the refractive angle caused by MS$_1$ is equaled to the incident angle in MS$_2$, that is $\theta_{tl1} = \theta_{il2}$. Meanwhile, the change rule of incident angle $|\theta_{il2}|$ is contrary to that of $|\theta_{tl2}|$, which means that the change rule of $|\theta_{tl1}|$ is contrary to that of $|\theta_{tl2}|$. As shown in Fig. 3(b), when light waves are normally incident on MS$_2$, then $|\theta_{tr2}| = 90°$, light waves are converted into the rapidly decaying surface wave and propagate along the surface of MS$_2$, and cannot arrive at the left side.

Owing to the distributions of phase delays of MS$_1$ and MS$_2$ are all symmetric about $y = 0$, it only needs to analyze the phase delays when $y > 0$. The theoretical model of AFL is shown in Fig. 4.

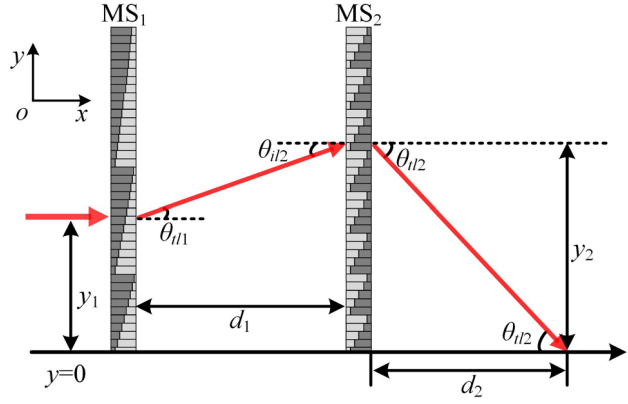

FIG. 4. Transmission path diagram when light waves are normally incident from the left side.

It can be seen from Fig. 4, light waves are focused on the right side through AFL, so the transmission directions of light waves need to meet the transmission path shown by the red solid line in Fig. 4, where $\theta_{tl1}$ and $\theta_{tl2}$ are the refractive angles caused by MS$_1$ and MS$_2$, respectively, when light waves are normally incident from the left side. The refractive angle $\theta_{tl2}$ is caused by the MS$_2$ when light waves are incident with $\theta_{il2}$, therefore, according to Eq. (1) and $\theta_{tl1} = \theta_{il2}$, $\theta_{tl1}$ and $\theta_{tl2}$ satisfy the following relation

$$\sin\theta_{tl2} = \sin\theta_{tl1} + \frac{1}{k}\frac{d\varphi_2}{dy} \tag{6}$$

where $d\varphi_2/dy$ is decided by $\varphi_2$, and $\varphi_2$ is the phase delay of MS$_2$. Owing to the structure of MS$_2$ coincides with that in Fig. 2(a), whose $\Delta l = -0.10\lambda$, and the phase delay of MS$_2$ is irrelevant to the incident angle, then according to Eqs. (2), (5) and substitute the relevant values, Eq. (6) can be expressed as

$$\sin\theta_{tl2} = \sin\theta_{tl1} - 1 \tag{7}$$

meanwhile, according to the geometric relationship, $d_1$, $d_2$, $y_1$, and $y_2$ satisfy the following relations

$$\tan\theta_{tl2} = -\frac{y_2}{d_2} \tag{8}$$

$$\tan\theta_{tl1} = \frac{y_2 - y_1}{d_1} \tag{9}$$

where $d_1$ is the distance between MS$_1$ and MS$_2$, $d_2$ is the focal length, $y_1$ and $y_2$ are the positions where light waves are incident on MS$_1$ and MS$_2$, respectively. Through Eqs. (8) and (9), $\theta_{tl1}$ satisfies the following relation

$$\tan\theta_{tl1} = \frac{-d_2 \tan\theta_{tl2} - y_1}{d_1} \tag{10}$$

By combining Eqs. (1) and (2), the relationship between $\varphi_1$ and $y_1$ can be obtained

$$\varphi_1 = k \int \sin\theta_{tl1} dy_1 \tag{11}$$

where $\varphi_1$ is the phase delay of MS$_1$. It can be seen from Eq. (11), when the phase delays of MS$_1$ and MS$_2$ interact with each other, the distribution of phase delay of MS$_1$ presents not generally periodic change but presents continuous change. If $d_1 = 5\lambda$ and $d_2 = 6\lambda$ are set, the distribution of phase delay of MS$_1$ and asymmetric focusing field when the operating wavelengths $\lambda = 5\mu m$ can be obtained by Eqs. (7), (10) and (11), as shown in Fig. 5.

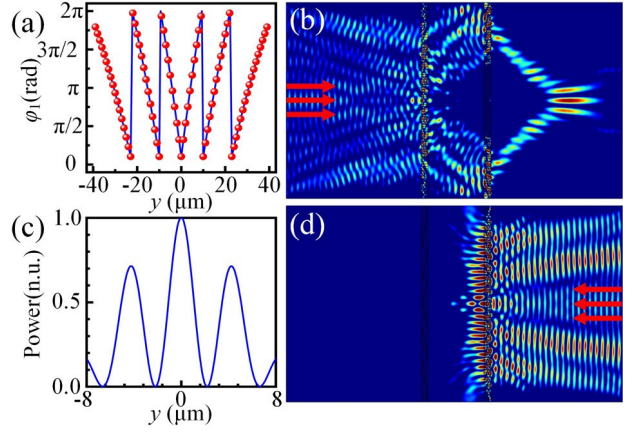

FIG. 5. Phase delay of $MS_1$ and diagram of asymmetric focusing: (a) distribution of phase delay of $MS_1$; (b) the field distribution when light waves are incident from the left side; (c) intensity curve of focal plane at focal point; (d) the field distribution when light waves are incident from the right side.

By combining the distribution of phase delay of $MS_1$ (see Fig. 5(a)) and Eqs. (2)-(5), $\Delta l$ of $MS_1$ at different positions in the y-axis can be obtained, which can be applied to design $MS_1$. It can be seen from Figs. 5(b) and 5(c), the focusing of the plane wave which is normally incident from the left side can be realized through AFL. However, its sidelobe is too large (the cause is analyzed and the result is optimized in Section 3.3), which is not conducive to practical application. As shown in Figs. 5(b) and 5(d), asymmetric transmission and focusing are realized simultaneously by AFL, which can be used to the more complex light field.

### 3.3 NZS asymmetric focusing realized by the optimized AFL

Then the cause of large sidelobe of asymmetric focusing in Fig. 5 is analyzed, and the structure of AFL is optimized to realize NZS asymmetric focusing.

It can be obtained from Eq. (9)

$$y_2 = y_1 + d_1 \tan\theta_{tl1} \qquad (12)$$

Owing to the structure of the lens is symmetric about the *x*-axis and for the sake of the convenience in theoretical analysis, here only need to analyze the upper part of AFL, which means $y_1 \geq 0$, that is $y_2 \geq d_1\tan\theta_{tl1}$. The minimal value of $y_2$ can be obtained when $y_1 = 0$, so $y_{2\min} \approx 3.16\lambda$. Meanwhile, the length of the upper part of $MS_2$ is $7.9\lambda$, which means that there is more than a third of the region in the middle of $MS_2$ not involved in the focusing. Therefore, the components of light waves involved in focusing are less and cause large sidelobe. To eliminate sidelobe for better focusing, the middle region of $MS_2$ has to be as far as possible to be involved in focusing. Therefore, the direction of the refractive angle caused by $MS_1$ should be contrary to that in Fig. 3 to make the light waves can pass the middle region of $MS_2$. To this end, the medium with refractive index $n_1$ and $n_2$, respectively, of $MS_1$ in Fig. 3 are switched positions. The theoretical models of focusing and transmission paths are as shown in Fig. 6.

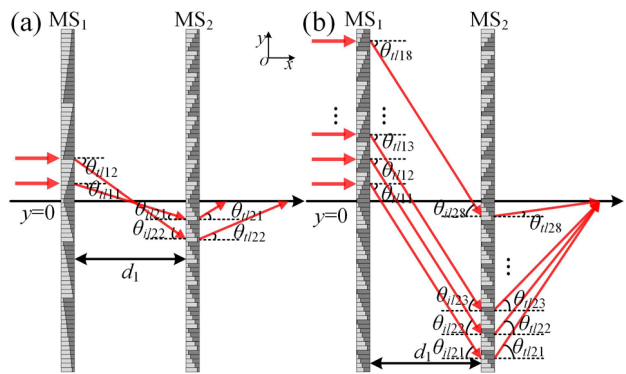

FIG. 6. The optimization of AFL: (a) the diagram of the transmission path of light waves which are normally incident from the left side when the phase delay of $MS_1$ presents continuous change but the refractive angle of $MS_1$ are contrary to that in Fig.

3, and the structure of MS$_2$ is consistent with that in Fig. 3; (b) the diagram of the transmission path of light waves which are normally incident from the left side when the phase delay of MS$_1$ presents periodic change and the phase delay of MS$_2$ presents gradient change.

In Fig. 6(a), the structure of MS$_2$ is consistent with that in Fig. 3, but the refractive angle caused by MS$_1$ is contrary to that in Fig. 3. Therefore, the change rule is contrary to that in Fig. 3 when light waves are normally incident from the left side, that is, $|\theta_{tl12}| > |\theta_{tl11}|$. Owing to the differences of MS$_1$ between Figs. 3 and 6 are only switching the position of mediums with different refractive indexes, a similar method can be used to derive the relationship satisfied by the lens in Fig. 6

$$-\sin\theta_{tl1} + \sin\theta_{tl2} = 1 \tag{13}$$

$$\tan\theta_{tl2} = \frac{y_2}{d_2} \tag{14}$$

$$\tan\theta_{tl1} = -\frac{y_2 + y_1}{d_1} \tag{15}$$

Therefore, by combining $|\theta_{tl12}| > |\theta_{tl11}|$ and Eq. (13), one can obtain $|\theta_{tl22}| < |\theta_{tl21}|$, that is, the refractive angle is not meet the distribution of focusing when the light waves are normally incident from the left side and most light waves cannot be involved in focusing, whose aberration is huge and cannot meet the actual demand.

To solve this problem, the phase delay distribution of MS$_1$ is changed to be periodic, while the phase delay of MS$_2$ no longer maintains the periodic change. If the phase delay of MS$_2$ changes continuously, the surface wave cannot be realized. For the simultaneous realizations of asymmetric transmission and focusing, the phase delay of $MS_2$ is set to gradient change, that is, the refractive angle caused by $MS_2$ presents gradient change, whose theoretical model is shown in Fig. 6(b). In Fig. 6(b), owing to the change rules of phase delays of $MS_1$ and $MS_2$, $|\theta_{tl1}|$ in each position at the y-axis is identical and $|\theta_{tl2}|$ varies directly as the distance from the x-axis, which can maintain the distribution rule of focusing. Meanwhile, the light waves are incident from the right side have to be transformed to surface wave through $MS_2$, and the relationship of $\theta_{tl1}$ and $\theta_{tl2}$ can be obtained by Eq. (13) (can also be obtained visually by Fig. 6(b))

$$\sin\theta_{tl2m} - \sin\theta_{tl1m} > \sin\theta_{tl2(m+1)} - \sin\theta_{tl1(m+1)} \geq 1 \quad (16)$$

where integer $m \in [1,7]$ represents the order number of refractive angles on different positions in $MS_2$. According to Eq. (6), $\theta_{tl1m}$ and $\theta_{tl2m}$ in this structure satisfy the following relation

$$\sin\theta_{tl2m} - \sin\theta_{tl1m} = \frac{1}{k}\frac{d\varphi_{2m}}{dy} \quad (17)$$

Therefore, the phase delay of $MS_2$ can be designed reasonably by combining Eqs. (16) and (17). Meanwhile, owing to $|\theta_{tl1}|$ in each position at the y-axis is identical, the parameter settings of metasurfaces can be determined by analyzing the value of $|\theta_{tl1}|$. If $|\theta_{tl1m}| > 55°$, according to Eq. (2), the phase delay of $MS_2$ cannot cover the entire range of $2\pi$ for the length restriction of $l$. If $|\theta_{tl1m}| < 45°$, according to Eq. (15), that is, $d_1 = -(y_1+y_2)/\tan\theta_{tl1}$, which means it needs larger $d_1$ to realize most regions of $MS_2$ are involved in focusing, but the larger $d_1$ limits the application of the optimized AFL. According to the above analysis, setting $|\theta_{tl1m}| = 50°$, while $d_1$ is consistent with that in Fig. 4, the phase delays of the dual-layer metasurfaces of the optimized AFL and field

distribution of NZS asymmetric focusing can be obtained, as shown in Fig. 7.

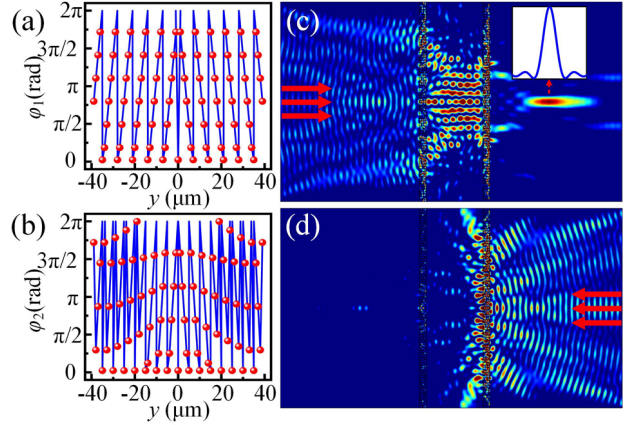

FIG. 7. The phase delays of the dual-layer metasurfaces of the optimized AFL and field diagram of NZS asymmetric focusing: (a) the distribution of phase delay of $MS_1$; (b) the distribution of phase delay of $MS_2$; (c) the field diagram when the light waves are incident from the left side, insert figure is intensity curve of focal plane at focal point; (d) the field diagram when the light waves are incident from the right side

It can be seen from Figs. 7(a) and 7(b), the distributions of phase delays of $MS_1$ and $MS_2$ present periodic and gradient changes, respectively. From Fig. 7(c), the NZS focusing can be realized by optimized AFL when light waves are normally incident from the left side, and the results are superior to Fig. 5(a). From Fig. 7(d), the light waves are still transferred to surface wave, and cannot pass the optimized AFL. Meanwhile, from the insert figure in Fig. 7(c), the sidelobe is far less than that in Fig. 5(c), which means the sidelobe of asymmetric focusing can be effectively reduced.

*3.4 The frequency ranges of NZS asymmetric focusing realized by the optimized AFL*

The frequency ranges, full width at half maximum (FWHM), and field diagram of NZS asymmetric focusing at different frequencies can be obtained by changing the incident frequencies, the results are shown in Fig. 8.

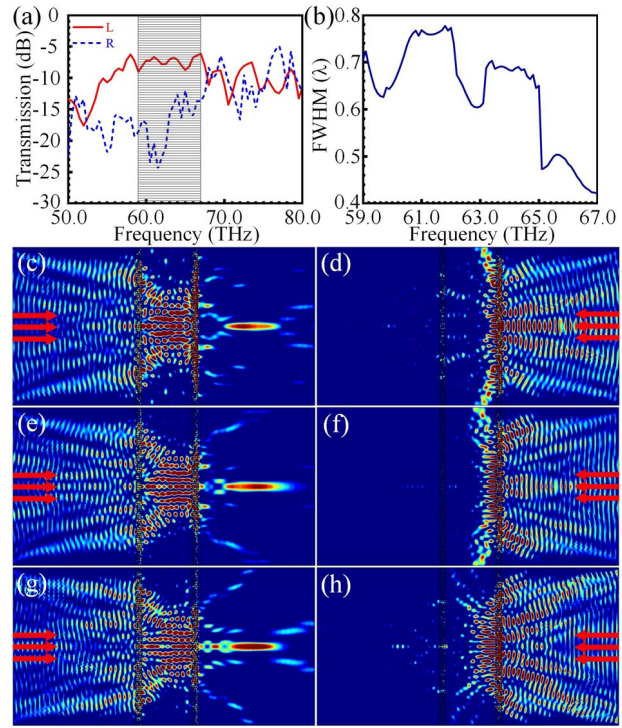

FIG. 8. The frequency ranges of NZS asymmetric focusing realized by the optimized AFL: (a) The transmission for plane waves are normally incident from the left and right sides, respectively, with different frequencies; (b) focusing FWHM of plane waves when they are normally incident from the left side with different frequencies; field diagram of NZS asymmetric focusing with different frequencies: (c)(d) 59.5 THz; (e)(f) 62.0 THz; (g)(h) 64.5THz.

Fig. 8(a) shows that the frequency ranges of NZS asymmetric focusing are $f \in$(59.0 THz, 67.0 THz). The NZS subwavelength focusing can be realized when plane waves are normally incident from the left side in these frequency ranges, as is depicted in Fig. 8(b). These results indicate that the optimized AFL can realize NZS asymmetric focusing in broadband, which has great practical application values. From

Figs. 8(c)-8(g), all lights with different frequencies (59.5THz, 62.0 THz, and 64.5 THz) can realize the NZS asymmetric focusing by the optimized AFL further indicates that the NZS asymmetric focusing possesses broadband characteristics.

## 4. Conclusions

We design an AFL with dual-layer metasurfaces based on generalized Snell's law for realizing optical asymmetric transmission and focusing simultaneously. Considering the initially designed AFL with drawbacks of the middle region of metasurface on the right side cannot be involved in focusing, as well as the sidelobe of optical asymmetric focusing is large, the design theory is optimized further, and the NZS asymmetric focusing on the subwavelength scale is realized. The NZS asymmetric focusing of the optimized AFL is effective in broadband, which has great application potential in micro/nano-optical devices.

## Acknowledgments

This work was supported by the National Natural Science Foundation of China (Grant Nos. 61405058, and 61605166), the Natural Science Foundation of Hunan Province (Grant Nos. 2017JJ2048, and 2018JJ3514), and the Fundamental Research Funds for the Central Universities (Grant No. 531118040112). The authors acknowledge Prof. J. Q. Liu for software sponsorship.